\documentclass[11pt,twoside]{article}
\usepackage{asp2010}
\resetcounters

\begin{document}

\title{Evolved stars with complex 
atmospheres -
the high spectral resolution, 
mid-IR view}
\author{N. Ryde$^1$, J. Lambert$^1$, M. J. Richter$^2$, E. Josselin$^3$, G. M. Harper$^4$, K. Eriksson$^5$, A. Boogert$^6$, C. DeWitt$^2$, T. Encrenaz$^7$, T. Greathouse$^8$, D. Jaffe$^9$, K. Kulas$^{10}$, M. McKelvey$^{10}$,J. Najita$^{11}$, W. Vacca$^6$ 
\affil{$^1$Department of Astronomy and Theoretical Physics, Lund Observatory, Lund University, Box 43, SE-221 00 Lund, Sweden}
\affil{$^2$Department of Physics, University of California at Davis, CA 95616, USA}
\affil{$^3$UPM, Universit\'e Montpellier II, Montpellier, France}
\affil{$^4$Astrophysics Research Group, Trinity College Dublin, Dublin 2, Ireland}
\affil{$^5$Department of Physics and Astronomy, Uppsala University, Box 516, SE-751 20 Uppsala, Sweden}
\affil{$^6$USRA-SOFIA Science Center, NASA Ames Research Center, Moffett Field, CA 94035, USA}
\affil{$^7$LESIA, Observatoire de Paris, CNRS, UPMC, Univ. Denis Diderot, 92195, Meudon, France}
\affil{$^8$Southwest Research Institute, Division 15, 6220 Culebra Road, San Antonio, TX 78228, USA}
\affil{$^9$The University of Texas, Austin, TX 78712, USA}
\affil{$^{10}$NASA Ames Research Center, Moffett Field, CA 94035, USA}
\affil{$^{11}$NOAO, 950 North Cherry Avenue, Tucson, AZ 85719, USA}}

\begin{abstract}
The physical structures of the outer atmospheres of red giants are not known. They are certainly complex and a range of recent observations are showing that we need to embrace to non-classical atmosphere models to interpret these regions. This region's properties is of importance, not the least, for the understanding of the mass-loss mechanism for these stars, which is not still understood. Here, we present observational constraints of the outer regions of red giants, based on mid-IR, high spectral resolution spectra. We also discuss possible non-LTE effects and highlight a new non-LTE code that will be used to analyse the spectra of these atmospheric layers. We conclude by mentioning our new SOFIA/EXES observations of red giants at $6\,\mu$m, where the vibration-rotation lines of water vapour can be detected and spectrally resolved for the first time.
\end{abstract}

\section{Observations}

\begin{figure}[h]
  \centering
	\includegraphics[angle=0, trim=0mm 0mm 0mm 0mm, clip, width=1.0\textwidth]{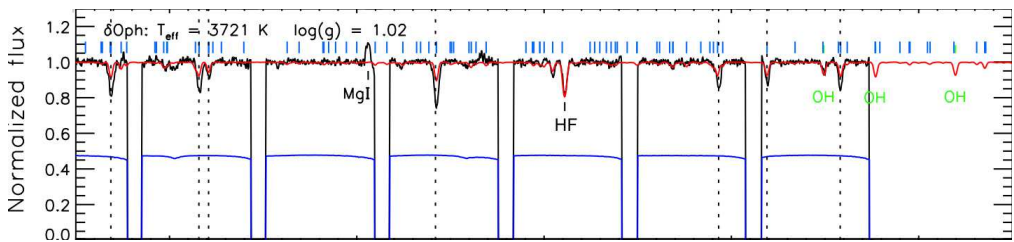}
	\caption{Example spectrum of $\delta$ Oph (M0.5 III) at $12\,\mu$m is shown in black. The red spectrum is the best synthetic spectrum for the star. In blue we show the telluric spectrum,. It is obvious that in this wavelength region there are no severe telluric absorption.} 
	\label{doph}
\end{figure}

We have presented new high spectral resolution $12\,\mu$m spectra of early-K to mid M giants of a range of effective temperatures $3400\,\mathrm{K} < T_\mathrm{eff} < 4900\,\mathrm{K}$ in Ryde et al, 2014, submitted. The spectra were recorded with the TEXES spectrograph \citep{texes} at the NASA-IRTF on Mauna Kea at an altitude of $4,205$\,m. An example spectrum of $\delta$ Oph (M0.5 III) is shown in Figure \ref{doph}. 

\section{Results} 

In all the spectra the synthetic profiles of the water lines are too weak. The emission lines of Mg and the HF lines are clearly detected and are modelled well with a pure photospheric model described in 
\citet{ryde:04_letter}, \citet{ryde:04_mg}, and \citet{sundqvist:08} for the Mg emission and in \citet{joensson:14b} and Herik J\"onsson's talk for the HF line in context of the galactic chemical evolution of fluorine. Also, other features such as the OH($v=2-2$, $3-3$) lines are well modelled.  We do not  detect, to within $1.5\,$km\,s$^{-1}$, any velocity shift between the detected spectral lines. Furthermore, we do not detect any emission of water whatsoever.
This means that at least the continuum and these lines are most likely formed in the photospheres of the stars and that no signatures of the more extended outer atmospheres are detected. 

To summarise, the results of our observations show deeper water lines that expected
in a large range of red giants, varying smoothly with spectral type. The stronger-than-expected lines are thus not object-specific, but a general feature of red giants, cf. also the results of Sloan et al. (2014, submitted).

In the spectra, we have detected water lines with a span of excitation energies. Thus, by assuming that they are formed in a region in the photosphere represented by a single temperature, we can calculate a characteristic excitation temperature,  from the relative lines strengths of these lines. We find that the derived temperatures are on the order of 500 K lower than that expected from LTE photospheric {\sc marcs} atmospheres \citep{marcs:08} of these stars. We further find that the water lines are very optically thick.


\section{Discussion}

The strong water lines are a general feature for all red giants, smoothly varying with effective temperature. 
Any model of the outer atmospheres of red giants need to accommodate this behaviour.
Most likely, these lines are photospheric in origin. The continuum is formed in the photosphere, the water lines are   
optically thick, and we do not detect any emission. These features are  difficult to explain with molecular layers around the stars (MOLsphere).

What causes the deeper photospheric lines? Below we list a number of suggestions. For a further discussion, see \citet{ryde:water0,ryde:water1,ryde:water2,ryde:water3}.
\begin{itemize}
\item non-LTE temperature structure (cf. \citet{short:03}: non-LTE in Fe)
\item non-LTE cooling (due to non-LTE in water, Lambert et al. 2014)
\item non-LTE H$_2$O line source function and line opacity (Lambert et al. 2014)
\item convective flows (3D red giants)
\item temperature bifurcations due to  molecular catastrophes
\item dynamic chromospheres
\item star spots
\end{itemize}

Here we will only discuss what we suggest to be the most likely, namely the importance of an non-LTE treatment of the line formation of the water lines. The reason for this is that when computing the critical density, it is clear that non-LTE effects need to be investigated at these atmospheric depths.

In order to investigate this further, we have developed an original code for the non-LTE calculation of massive molecular models, such as that for the water molecule. There are more than 1000 levels to consider up to an excitation energy of $5000\,$cm$^{-1}$, with more than $15\,000$ radiative and $350\,000$ collisional rates. This is not possible to handle with other non-LTE codes available, and is too large for a classical radiative-transfer approach. The method, fully parallelized in an MPI code, is explained in detail in \citet{julien:11,julien:12,julien:13}, Lambert et al. (2014, submitted) and the poster presented at this conference (Lambert et al.). What they find is that there are large departure coefficients and that the spectrum is affected. More importantly, the water cooling of the outer photosphere is affected. The detailed response of these affects will be presented in a forthcoming paper. Thus, we find that these non-equilibrium level populations will affect the cooling, line source function, and line opacities. 
It is clear that non-LTE will be important to consider. This might be important for all outer atmospheres of red giants including AGB stars. The non-LTE tools now exist and we will investigate these effects for AGB stars.




\subsection{SOFIA - the Stratospheric Observatory For Infrared Astronomy}

Up to now we have only been investigating the rotational water lines at $12\,\mu$m. We have, however, also started a program to investigate the vibration-rotation ($\nu_2$) lines at $6.5\,\mu$m at high spectral resolution ($R\sim 85,000$) with EXES \citep[see][]{sofia:exes:SV} on SOFIA. This is the first time a star has been investigated at high spectral resolution at these wavelengths, which are totally opaque from the ground. Our first observations were performed on the first commissioning flight of SOFIA/EXES on April, 9 2014, and will be presented in a forthcoming paper.
 

EXES fills a gap in the high spectral resolution, mid-infrared wavelength region ($4.5-28\,\mu$m).
Flying in the stratosphere, the SOFIA observatory flies above $99\%$ of the telluric water vapour, which means that apart from much less contamination of telluric water-vapour lines, the totally opaque (from the ground) regions between the M and N bands at $5.5-7.0\,\mu$m are also observable.

\section{Conclusions}

Red giants outer atmospheres are complex. Non-classical models are definitely needed, be it in context of a  MOLsphere or a non-classical photosphere scenario. There are contradictory views that have to be unified for a global view of these regions.

We have detected strong water lines in absorption in the N band for a wide range of giants, from early K to mid-M, that is 3500-4300 K. Our IRTF/TEXES spectra at $12\,\mu$m an SOFIA/EXES spectra at $6.5\,\mu$m provide strong constraints on any model attempt of these atmspheres.

We argue that the lines detected are probably photospheric in origin. The excitation temperature are found to be cooler than expected from {\sc marcs} models. This could indicate a physically cooler outer temperature structure due to non-LTE  water cooling, or a non-LTE line formation of the water lines as such.




\end{document}